\magnification=1200
\voffset=-1.5truecm\hsize=16.5truecm    \vsize=24.truecm
\baselineskip=14pt plus0.1pt minus0.1pt \parindent=12pt

\lineskip=4pt\lineskiplimit=0.1pt      \parskip=0.1pt plus1pt 
\def\st{\scriptstyle} 
 

\input eplain


\beginpackages
\usepackage{graphicx,color}
\usepackage{url}
\endpackages

 
\input blackdvi   

 
\let\a=\alpha   \let\d=\delta \let\e=\varepsilon 
\let\f=\varphi        
\let\m=\mu      \let\p=\pi  
\let\r=\rho \let\s=\sigma   
   
\let\D=\Delta  \let\G=\Gamma

 
\global\newcount\numsec\global\newcount\numfor 
\gdef\profonditastruttura{\dp\strutbox} 
\def\senondefinito#1{\expandafter\ifx\csname#1\endcsname\relax} 
\def\SIA #1,#2,#3 {\senondefinito{#1#2} 
\expandafter\xdef\csname #1#2\endcsname{#3} \else 
\write16{???? il simbolo #2 e' gia' stato definito !!!!} \fi} 
\def\etichetta(#1){(\veroparagrafo.\veraformula) 
\SIA e,#1,(\veroparagrafo.\veraformula) 
 \global\advance\numfor by 1 
 \write15{@def@equ(#1){\equ(#1)} \%:: ha simbolo= #1 } 
 \write16{ EQ \equ(#1) ha simbolo #1 }} 
\def\etichettaa(#1){(A\veroparagrafo.\veraformula) 
 \SIA e,#1,(A\veroparagrafo.\veraformula) 
 \global\advance\numfor by 1\write16{ EQ \equ(#1) ha simbolo #1 }} 
\def\BOZZA{\def\alato(##1){ 
 {\vtop to \profonditastruttura{\baselineskip 
 \profonditastruttura\vss 
 \rlap{\kern-\hsize\kern-1.2truecm{$\scriptstyle##1$}}}}}} 
\def\alato(#1){} 
\def\veroparagrafo{\number\numsec}\def\veraformula{\number\numfor} 
\def\Eq(#1){\eqno{\etichetta(#1)\alato(#1)}} 
\def\eq(#1){\etichetta(#1)\alato(#1)} 
\def\Eqa(#1){\eqno{\etichettaa(#1)\alato(#1)}} 
\def\eqa(#1){\etichettaa(#1)\alato(#1)} 
\def\equ(#1){\senondefinito{e#1}$\clubsuit$#1\else\csname e#1\endcsname\fi}

 
\def\bb{\hbox{\vrule height0.4pt width0.4pt depth0.pt}}\newdimen\u 
\def\pp #1 #2 {\rlap{\kern#1\u\raise#2\u\bb}} 
 
\def\ins #1 #2 #3 {\rlap{\kern#1\u\raise#2\u\hbox{$#3$}}}

\def\pallina{{\kern-0.4mm\raise-0.02cm\hbox{$\scriptscriptstyle\bullet$}}} 
\def\palla{{\kern-0.6mm\raise-0.04cm\hbox{$\scriptstyle\bullet$}}} 
\def\pallona{{\kern-0.7mm\raise-0.06cm\hbox{$\displaystyle\bullet$}}} 
\def\bull{\vrule height .9ex width .8ex depth -.1ex } 
 
\def\data{\number\day/\ifcase\month\or gennaio \or febbraio \or marzo \or 
aprile \or maggio \or giugno \or luglio \or agosto \or settembre 
\or ottobre \or novembre \or dicembre \fi/\number\year} 
 
 
\newcount\pgn \pgn=1 
\def\foglio{\number\numsec:\number\pgn 
\global\advance\pgn by 1} 
\def\foglioa{a\number\numsec:\number\pgn 
\global\advance\pgn by 1} 
 
\footline={\rlap{\hbox{\copy200}\ $\st[\number\pageno]$}\hss\tenrm \foglio\hss} 
 
 
\def\sqr#1#2{{\vcenter{\vbox{\hrule height.#2pt 
\hbox{\vrule width.#2pt height#1pt \kern#1pt 
\vrule width.#2pt}\hrule height.#2pt}}}}

 \def\\{\noindent}

\let\dpr=\partial

\def\ul#1{\underline#1}

\def\tende#1{\vtop{\ialign{##\crcr\rightarrowfill\crcr 
              \noalign{\kern-1pt\nointerlineskip} 
              \hskip3.pt${\scriptstyle #1}$\hskip3.pt\crcr}}} 
\def\otto{{\kern-1.truept\leftarrow\kern-5.truept\to\kern-1.truept}} 
 
\font\smfnt=cmr8 scaled\magstep0 
\font\myfonta=msbm10 scaled \magstep0 
\def\math#1{\hbox{\myfonta #1}}


\vglue.5truecm
{\centerline{\bf ON A FINITE RANGE DECOMPOSITION OF THE RESOLVENT } 
{\centerline{\bf OF A FRACTIONAL POWER OF THE LAPLACIAN }
{\centerline{\bf II. THE TORUS}

\vglue1cm
{\centerline{\bf P. K. Mitter}
\vglue1cm 
{\centerline{\smfnt Laboratoire Charles Coulomb}}
{\centerline{\smfnt CNRS-Universit\'e Montpellier- UMR5221}} 
{\centerline{\smfnt Place E. Bataillon, Case 070,  
34095 Montpellier Cedex 05 France}} 
\vglue0.5cm
{\centerline{\smfnt e-mail: Pronob.Mitter@umontpellier.fr}}

\vglue1cm
\\{\bf Abstract}: In  previous papers, [1], [2], we proved  the existence as well as regularity of a
finite range decomposition for the resolvent $G_{\a} (x-y,m^2) =
((-\D)^{\a\over 2} + m^{2})^{-1} (x-y) $,
for $0<\a<2$ and all real $m$, in the lattice ${\math Z}^{d}$  for dimension $d\ge 2$.  In this
paper, which is a continuation of the previous one, we extend those
results by proving the existence as well as regularity of a 
finite range decomposition for the same resolvent
but now on the lattice torus ${\math Z}^{d}/L^{N+1}{\math Z}^{d} $ 
for $d\ge 2$ provided  $m\neq 0$ and $0<\a<2$. We also prove
differentiability and uniform 
continuity properties with respect to the resolvent parameter $m^{2}$.
Here $L$ is any odd
positive integer and $N\ge 2$ is any positive integer.

\vglue0.5cm
\\{\bf 1. Introduction} 
\numsec=1\numfor=1
\vglue0.5cm
In  previous papers [1] and [2]   we proved  the existence as well as regularity of a
finite range decomposition for the resolvent 

$$G_{\a} (x-y,m^2) =((-\D)^{\a\over 2} + m^{2})^{-1} (x-y) \Eq(1.1)$$

\\for $0<\a<2$ and all real $m$, in the lattice ${\math Z}^{d}$  for
dimension $d\ge 2$.  The definition and properties of a finite range
decomposition were given in [1, 2]. The reference [2] incorporates the
content of the published version [1] together with its erratum.
and will thus be convenient to refer to.
The main result  is Theorem 1.1 of [1], restated in [2].
In this paper, we will prove for all $\a$ in the interval  $0<\a<2$ the existence and regularity of a finite
range decomposition of a periodic version of \equ(1.1) on the torus 
${\math Z}^{d}/L^{N+1}{\math Z}^{d} $. This is the content of Theorem
1.1 below. Continuity and differentiability properties in $m^{2}$ are
given in Theorem 1.2 below for $\a$ in the interval $1< \a
<2$. 
We
emphasise that Theorems 1.1 and 1.2 are valid only when $m\neq
0$. Results for finite range decompositions of general families of massless models on the discrete
torus are given in [9].

\vglue.03cm
The
resolvent \equ(1.1) arises as the covariance of the Gaussian measure
underlying various statistical/field theoretic systems with long range
interactions (see [2]). For $d=2, 3$ the upper critical dimension for
those systems is $d_c= 2\a$. Thus for  $\a$ in the above interval we can
arrange for the system to be below the upper critical dimension, and
this is where non-trivial critical phenomena for long range systems
are expected.
 
\vglue0.3cm

For $\a=2$ the resolvent in \equ(1.1) is that of a standard massive
Laplacian. A finite range decomposition on the lattice  ${\math Z}^{d}$  was obtained in [3]. 
In this case also the methods of this paper can  be applied
for obtaining a finite range decomposition on the  lattice torus 
starting from the work in [3].  A finite range decomposition for the
resolvent of a massive Laplacian on the
lattice torus was obtained earlier in [10] using the results of [4].


\vglue0.3cm

This paper is a companion to the earlier papers [1, 2]. We will use freely the notations and
results, especially Theorem 1.1, Corollary 1.2 and Proposition 2.1, of
these references. 
However, for the convenience of the reader, before
embarking on the proofs of Theorems 1.1 and Theorem 1.2 we will
give some indications of the strategy to be followed making use of
results from earlier references. 


\vglue0.3cm

A survey of earlier results and references  together with motivation was given
in [1] and we will not repeat them here. We simply remind the reader
that  one of the main applications of finite range decompositions is
in rigorous Renormalisation Group analysis of statistical/field
theoretic systems near and at the critical point of second order phase transitions.
The lattice acts as an ultraviolet cutoff but we also need a finite volume
cutoff and then later take the infinite volume limit. In the finite
volume theory it is desirable to preserve translation invariance. One
convenient way of doing that is putting the theory on a torus of
finite period  which is the edge length of the fundamental domain in
the shape of a square, cube or hypercube.  The goal of extending the finite range
decomposition of the resolvent \equ(1.1) given in [1, 2]  (together with regularity properties) to the torus
 is achieved in this paper. 

We should point out that a different way of
 achieving the same goal  has been given in [6]
 using estimates from [5]. This is an essential ingredient in [6] where  critical exponents below the
 critical dimension $d_c=2\a$ have been studied for the 
 $n$-component $\f^{4}$ model with long range interactions  in the
 regime where $\e=d_c-d= 2\a-d>0$ is held sufficiently small.
Our method and results for extending the finite range decomposition to the torus however
differ from that in [6] and therefore we are providing them. It
relies on the bounds of Theorem 1.1 and Corollary 1.2 of [1] and [2] together with
Fourier analysis on the discretized torus using the discrete Fourier
 transform and estimates on Fourier
coefficients. These are obtained using estimates in [3] and the
spectral decomposition given in Proposition 2.1 of [1, 2].  The mass
derivatives of the functions appearing in the finite range decomposition
are estimated very simply using the spectral
decomposition and estimates given in [1], and [2]. Continuity
 results are consequences. Mass derivative/continuity bounds are known
 to be useful in the study of critical exponents [6].  A
comparison with the bounds given in [6] is given later in Remarks 3
 and 5 below. 

\vglue 0.3cm
\\{\bf Definitions}
\vglue0.3cm
\\Let $L=3^{p}$, $p\ge 2$, $\e_{j}=L^{-j}$, $j\ge 0$,  $0 < \a <2 $
and $d\ge 2$.  Let $N\ge 2$ be any positive integer. 
Let $T_{N+1} (=T_{N+1}^{d})$  denote the torus ${\math Z}^{d}/L^{N+1}{\math Z}^{d}$ of
edge length $L^{N+1}$. 
The fundamental cube 
$Q_{N+1}= [-{L^{N+1}\over 2}, {L^{N+1}\over 2}]^{d}\cap  {\math
Z}^{d}$ has the property that every point of ${\math Z}^{d}$ 
has a unique translate  with respect to $Q_{N+1}$. 
The volume of the fundamental cube is $| Q_{N+1}| =L^{(N+1)d}$. 
Functions on the torus are periodic functions. Integration
(summation) on the torus is defined as usual as integration over the
fundamental cube.  Moreover  
we define $L^{1}(T_{N+1})= L^{1}(Q_{N+1}) $. If $X\subset {\math
Z}^{d}$ then $L^{1}(X)$ is the space of summable functions on $X$.

\vglue0.3cm

In the following
we often speak of {\it periodizing}  a function. 
Let $f: {\math Z}^{d}\rightarrow {\math R}$. We say that $f$ has a
periodization $f_{T_{N+1}}$ with period $L^{N+1}$ with $N$ any positive
integer if for $\forall x\in {\math Z}^{d}$ the sum

$$f_{T_{N+1}}(x)= \sum_{y\in L^{N+1}{\math Z}^{d}} f(x+ y) \Eq(1.011) $$

\\exists. If $f\in L^{1}({\math Z}^{d}) $  then the sum converges absolutely in $L^{1}(Q_{N+1})$ 
and defines 
$f_{T_{N+1}}$ 
as a function in
$L^{1}(T_{N+1})$. For analogous onsiderations in the continuum see
e.g. Stein and Weisz, Chapter 7,  in [7].

\vglue0.3cm

\\Finally we note that we shall often
employ  continuum integral notations for lattice sums.
The Lebesgue measure in $(\e_n{\math Z})^{d}$ is the counting
meaure times $\e_{n}^{d}$.

\vglue0.3cm
\\All objects in the following Theorem 1.1  will be defined and introduced
below immediately after the statement of the theorem.

\vglue0.3cm

\\{\bf Theorem 1.1}

\\Let $0<\a<2$,  $d\ge 2$ and $N\ge 2$. Let $\e_j=L^{-j},\> \forall
j\ge 0$ and
let $m\neq 0$. Then the positive definite function $G_{\a}(x-y,m^2)$ on ${\math Z}^{d}$
has a periodized version $G_{\a, T_{N+1}}(x-y,m^2)$
which is a function in  $L^{1}(T_{N+1})$.  Moreover for all $m\neq 0$  we have the
the following finite range decomposition:

$$G_{\a, T_{N+1}}(x-y,m^2)= \sum_{j= 0}^{N-1} \> L^{-2j[\f]} \> 
\G_{j,\a}({x-y\over L^{j}}, L^{j\a}m^{2}) +  L^{-2N[\f]} {\cal G}_{N,\a,  T_{N+1}}  ({x-y\over L^{N}}, L^{N\a}m^{2})   \Eq(1.2)  $$

\\where 

$$[\f]= {d-\a \over 2}  \Eq(1.21)$$ 

\\and the positive definite functions $\G_{j,\a}(\cdot, m^{2})$,
defined on
$(\e_j{\math Z})^{d}$, which appear  in the sum  are those in
Theorem 1.1 of [1] and [2].  
The function ${\cal G}_{N,\a,
T_{N+1}}$  which did not appear in [1], {2] will be defined later at the end of this theorem.
The functions $\G_{j,\a}(\cdot, m^{2})$  have finite range $L$ and satisfy the bounds
stated in [1] and [2]:

\vglue0.3cm

\\For all $j\ge 2$ and $0\le q\le j$, and all $p\ge 0$,

$$|| \dpr _{\e_j} ^{p} \G_{j,\a} (\cdot, m^{2}) ||_{L^{\infty}
((\e_q{\math Z})^{d})} \le c_{L, p,\a} (1+ m^{2})^{-2} . \Eq(1.10) $$

\\For $j=0,1$ and $0\le q\le j$ we have the bound 

$$|| \dpr _{\e_j} ^{p} \G_{j,\a} (\cdot, m^{2}) ||_{L^{\infty}
((\e_q{\math Z})^{d})} \le c_{L, p,\a} (1+ m^{2})^{-1} . \Eq(1.101) $$

\\In the above $\dpr _{\e_j} = \dpr _{\e_j, e_k} ,\> k=1,..,d$ is a forward lattice partial derivative with
increment $\e_j$  and in any particular direction $e_k$ in the lattice
$(\e_j {\math Z}) ^{d} $.  Moreover
$\dpr_{\e_j} ^{p} $ is a multi-derivative of order $p$ defined as in the
continuum  but now with lattice forward derivatives. $e_1,....,e_d$
are unit vectors which give the orientation of ${\math R}^{d}$ as well
as the orientation of all embedded lattices $(\e_j {\math Z}) ^{d}
\subset {\math R}^{d}$. By construction the lattices are nested in an
obvious way. The constant $c_{L, p,\a}$ depends on $L, p, \a$. It depends
implicitly on the dimension $d$. 
\vglue0.3cm  

\\The functions on ${\math Z}^{d}$  

$$\tilde\G_{j,\a} (x, m^{2})= L^{-2j[\f]} \> \G_{j,\a} ({x\over L^{j}}, L^{j\a}m^{2}) \Eq(1.3)$$ 

\\have finite range $L^{j+1}$

$$\tilde\G_{j,\a} (x, m^{2}) =0\> :\> |x|\ge L^{j+1} \Eq(1.4) $$

\\and therefore  for $0\le j \le N-1$ are functions on
$T_{N+1}$. Their periodization give back the functions themselves.
They satisfy the regularity bounds of Corollary 1.2 of [1]  and [2]: 

\\for $j\ge 2$,

$$ ||\dpr_{{\math Z} ^{d}}^{p} \tilde \G_{j,\a} (\cdot,
m^{2})||_{L^{\infty} ({\math Z}^{d})}  \le
c_{L, p,\a} (1+ L^{j\a} m^{2})^{-2} L^{-(2j[\f] +pj)}    \Eq(1.14)$$ 

\\and for $j=0,1$ 

 $$ ||\dpr_{{\math Z} ^{d}}^{p} \tilde \G_{j,\a} (\cdot,
m^{2})||_{L^{\infty} ({\math Z}^{d})}  \le
c_{L, p,\a} (1+ L^{j\a} m^{2})^{-1} L^{-(2j[\f] +pj)}   .  \Eq(1.141)$$

\\For all $N\ge 2$ the function 

$$ \tilde{\cal G}_{N,\a,  T_{N+1}}  (x, m^{2})=L^{-2N[\f]}
 {\cal G}_{N,\a, T_{N+1}}  ({x\over L^{N}}, L^{N\a}m^{2}) \Eq(1.6)$$

\\is in $L^{1}(T_{N+1})$ and satisfies for all $m\neq 0$

$$|\dpr_{{\math Z}^{d}} ^{p}\> \tilde{\cal G}_{N,\a,T_{N+1}}(x,m^{2})|
\le c_{L,\a, p}  L^{-2N\a} m^{-4} \>  L^{-(2N[\f] +pN)} . 
\Eq(1.7)$$

\\For $m^{2} \ge {1\over\sqrt C}
L^{-N\a}$ where $C$ is any positive constant  independent of $N$,
and all integers $p\ge 0$, we therefore get the bound

$$||\dpr_{ {\math Z}^{d} }^{p} \tilde{\cal G}_{N,\a,  T_{N+1}}
(\cdot, m^{2})||_{L^{\infty} (T_{N+1})}  \le
c_{L, p,\a} \>  L^{-(2N[\f] +pN)}   \Eq(1.77)     $$
\\where the constant $c_{L, p,\a}$ depends on $C, L, \> p,\> \a$ but is independent of $N$, $m$.

\vglue0.3cm
\\{\bf A guide to Theorem 1.1} 

\vglue0.3cm

\\We  recall for the benefit of the reader the basic objects
introduced above. The functions
$$\G_{j,\a} (\cdot,\>m^{2} ) :\>(\e_{j} {\math Z} )^{d} \rightarrow {\math R}$$

\\are defined by

$$ \G_{j,\a} (\cdot,\>m^{2} )= \int_{0}^{\infty} ds\>\r_{\a}(s,\>
m^{2})\> \G_{j} (\cdot,\>s ) .  \Eq(1.151)$$

\\where $\G_{j} (\cdot,\>s )$ is the rescaled fluctuation covariance
in the finite range decomposition of the resolvent of the standard 
Laplacian (see [3]]) and $\r_{\a}(s,\>m^{2})$ is the
spectral function given by Proposition 2.1 in [1, 2]:

$$\r_{\a} (s, m^{2}) = {\sin\p\a/2\over \p} \> {s^{\a/2}\over s^{\a}
+m^{4} + 2 m^{2} s^{\a/2} \cos\p\a/2} \> . \Eq(1.152)$$

\\This latter function has bounds given in [1, 2] and these can be found
again in Section 3 of the present paper. These bounds together with the bounds
on $\G_{j} (\cdot,\>s )$ of [3] were used in [1, 2] to provide the
bounds on the fluctuation covariances $\tilde\G_{j,\a} (\cdot, m^{2})$
in Theorem 1.1 above. 

\vglue0.3cm

\\The function $\tilde{\cal G}_{N,\a,  T_{N+1}}
(\cdot, m^{2})$ is the periodization of a function $\tilde{\cal
G}_{N,\a}  (\cdot, m^{2})$ on ${\math Z}^{d}$ which is shown to be in
$L^{1}({\math Z}^{d})$. The latter function is
 the unrescaled version of the function
${\cal G}_{N,\a}  (\cdot, m^{2})$ on $(\e_{N} {\math Z})^{d}$ given by

$${\cal G}_{N,\a}  (\cdot, m^{2})= \int_{0}^{\infty} ds\>\r_{\a}(s,\>
m^{2})\> {\cal G}_{N} (\cdot,\>s ) . \Eq(1.153) $$

\\In [3] the finite range decomposition for the standard Laplacian on
${\math Z}^{d}$  was given for an arbitrary but finite number of terms
together with an explicit formula for the remainder which is
${\cal G}_{N} (\cdot,\>s ) $. This formula for the remainder is given
and used later in Section 2 in the course of proving the statements 
about $ \tilde{\cal G}_{N,\a,  T_{N+1}}  (\cdot, m^{2})  $ in Theorem 1.1 above. Rescalings
were performed in [3] which is why ${\cal G}_{N} (\cdot,\>s ) $ is a
function on the lattice $(\e_{N} {\math Z})^{d}$.


\vglue0.3cm
\\{\bf Remark 1: Scale independence of constants}

\vglue0.3cm

\\As in [1] (erratum) and [3], one can get rid of the scale dependence of constants by 
coarse graining on a larger scale $L'=L^{r}$ with $r$ a large positive
integer and holding $L$ fixed. The finite range expansion can be
rewritten by summing the fluctuation covariances and the remainder
over the intermediate scales. The fluctuation
covariances on the coarser scale $L'$  are defined by

$$\tilde \G'_{j,\a} (\cdot,m^{2})= \sum_{l=0}^{r-1} \tilde \G_{l+jr,\a} (\cdot,
m^{2})  . \Eq(1.1411)$$ 

\\We now get the coarse scale finite range decomposition

$$ G_{\a} (\cdot , m^{2}) = \sum_{j\ge 0} \tilde\G'_{j,\a} (\cdot,m^{2})  \Eq(1.1412) $$

\\with

$$\tilde \G'_{j,\a} (x-y,m^{2}) =0,\> |x-y|\ge (L')^{j+1} .\Eq(1.142)$$

\\The bounds on the coarse scale
fluctuation covariances and the coarse scale remainder remain unchanged with
new constants which are independent of the coarse scale $L'$. This is
explained in [1] as well as in 
[2] in the paragraph on coarse graining which follows Corollary 1.2
and is proved in  Appendix A of [2]. 

\vglue0.3cm
\\{\bf Remark 2}:
\vglue0.3cm
\\The function $\tilde{\cal G}_{N,\a,  T_{N+1}} $ on the torus,
introduced earlier, can also be viewed as the sum of all the functions $\tilde\G_{j,\a} $ for
$j\ge N$. In [6] the function was estimated as the   
the sum of estimates of the summands. Instead we estimate this
function directly using its explicit representation together with
Fourier analysis and estimates on the discrete Fourier transform. 

\vglue0.3cm
\\{\bf Remark 3}:
\\The bounds \equ(1.14) and \equ(1.141) on the fluctuation covariances
differ from those given in Proposition 10.1 of [7] as was noted
earlier in [2] and [3].  In particular the $(1+ L^{j\a} m^{2})^{-1}$
term in the bounds occur only for $j=0, 1$ terms and not for $j\ge 2$
in contrast to that in Proposition 10.1 of [6] where it occurs in the
bounds for all $j$.
However the bound \equ(1.7) agrees with the
relevant bound in Proposition 10.1 of [6] once one takes account of
the scale dimension $[\f]$ of the Gaussian field $\f$  (which is $2[\f] =d-\a$).

\vglue0.3cm
\\We have the following continuity and differentiability
properties in $m^{2}$ of the functions appearing in the finite range
decomposition \equ(1.2) of Theorem 1.1. They are given in Theorem 1.2 for $\a$
restricted to the  interval $1< \a<2$  and $d\ge 2$ for reasons explained in the introduction.

\vglue0.3cm

\\{\bf Theorem 1.2}
\vglue0.3cm

\vglue0.3cm

\\1. {\it Differentiability of fluctuation covariances} : Let $1< \a<2$ and $d\ge 2$.
For all $m^{2} >0$ and all $j\ge 1$,  the functions $\tilde\G_{j,\a}(\cdot, m^{2})$ 
are differentiable functions of $m^{2}$ and the derivatives
satisfy the bounds:
 
\vglue0.3cm
\\For $1< \a<2$ and all integers $p\ge 0$

$$||{\dpr \over \dpr m^{2}} \dpr_{ {\math Z}^{d} }^{p} \tilde\G_{j,\a}(\cdot, m^{2})
||_{L^{\infty} ({\math Z}^{d})} \le 
c_{L, \a, p} L^{-pj} L^{-j(d-2)} (m^{2})^{-2(1-{1\over \a})} . \Eq(1.102)$$

\vglue0.3cm

\\{\it Uniform Continuity}: As a consequence for all $m^{2}>0$,
$\tilde\G_{j,\a}(\cdot, m^{2})$ is a uniformly continuous function of
$m^{2}$. For all $m_{i} ^{2} > 0:\> i=1,2$ we have the following uniform
 bounds:

\vglue0.3cm

\\For $1<\a<2$ and all integers $p\ge 0$

$$||\dpr_{ {\math Z}^{d} }^{p} \tilde\G_{j,\a}(\cdot, m_{1}^{2}) -
\dpr_{ {\math Z}^{d} }^{p} \tilde\G_{j,\a}(\cdot, m_{2}^{2})||_{L^{\infty} ({\math Z}^{d})} 
\le c_{L, \a, p} L^{-j(d-2)} L^{-pj}
 \bigl|(m_{1}^{2}) ^{({2-\a\over \a})} -  (m_{2}^{2}) ^{({2-\a\over
 \a})} \bigr| . \Eq(1.103)$$

\vglue0.3cm

\\The constants $c_{L, \a, p} $ in \eq(1.101) and \equ(1.103) are independent of $j,m_{1}^{2}, m_{2}^{2}$.

\vglue0.3cm
\\2. {\it Differentiability of }  $\tilde{\cal G}_{N,\a,  T_{N+1}}
(\cdot, m^{2})$ : For all $m\neq 0$ and all integers $p\ge 0$

$$||{\dpr\over \dpr m^{2}} \dpr_{ {\math Z}^{d} }^{p} \tilde{\cal G}_{N,\a,  T_{N+1}}  (\cdot, m^{2})||_{L^{\infty} ({\math Z}^{d})}
\le   c_{L,\a, p}\> L^{-pN}  L^{-(N+1)d}   (m^{2})^{-2}  \Eq(1.104)$$

\\where the constant $c_{L, \a, p} $ is independent of $N$.
As a consequence we have the following

\vglue0.3cm

\\{\it Uniform continuity of } $\tilde{\cal G}_{N,\a,  T_{N+1}}  (\cdot, m^{2})$
\vglue0.3cm

\\For all $m_i\neq 0,\> i=1,2$ and all integers $p\ge 0$

$$\eqalign{&|| \dpr_{ {\math Z}^{d} }^{p} \tilde{\cal G}_{N,\a,  T_{N+1}}  (\cdot, m_{1}^{2}) -
\dpr_{ {\math Z}^{d} }^{p} \tilde{\cal G}_{N,\a,  T_{N+1}}  (\cdot, m_{2}^{2})||_{L^{\infty}({\math Z}^{d})} \cr
&\le c_{L, \a, p} \>L^{-pN}L^{-(N+1)d}\> m_{1}^{-2}  m_{2}^{-2}
|m_{1}^{2} -m_{2}^{2}|  .}  \Eq(1.105)$$

\vglue0.3cm

\\{\bf Remark 4: Scale independence of constants in mass derivative estimates}:  

\vglue0.3cm

\\For $d\ge 3$ ,$\forall p\ge 0$ as well as for $d=2$ , $\forall p\ge 1$ we can get rid of the scale dependence of the constants
in the bounds \equ(1.102) by passing to a coarser scale $L'$ as in
Remark 1 above with $L'=L^{r}$ with $L\ge 2$ fixed and $r$ a large positive
integer. The mass  differentiability bound on the coarse scale
fluctation covariance $\tilde\Gamma'_{j,\a}(\cdot, m^{2}) $ is then obtained
following Appendix A of [2] but now using the bound \equ(1.102) in
intermediate steps. Then the bound \equ(1.102) continues to hold for
$\tilde\Gamma'$ with a new constant $c'_{L,\a, p}$ independent of $L'$.
These statements are proved in Appendix A of the present paper.

\vglue0.3cm

\\For $d=2$ with $p=0$ the bound \equ(1.102) cannot be employed directly and we
have to proceed otherwise. We coarse grain the fluctuation covariances in [3] thus
producing a $\log L'$ dependence in bounds as was first done in [4].
This produces $\tilde\Gamma'_{j,\a}(\cdot, m^{2}) $ 
by the steps in Section 3 of [1, 2]. The bound \equ(1.101) now holds for $\tilde\Gamma'_{j,\a}(\cdot, m^{2}) $ 
with a new constant $c'_{L,\a, p} \log L'$ where $c'_{L,\a, p}$ is
independent of $L'$.
 
\vglue0.3cm

\\{\bf Remark 5}:

\\Note that the mass derivative bound given in \equ(1.102) agrees
after coarse graining (see Remark 4 above) with
that given in Proposition 10.1 of [6] for $d=3$ (once one has taken
 account of the definition of $\e=2\a -d$ which figures in the bounds in [6]).
For $d=2$ with $p\ge 1$ we have no logarthmic dependence either in the
 scale or the mass (see Remark 4 above) in contrast to Proposition
 10.1 of [6]. 
For $d=2$ with $p=0$ we have a
logarithmic scale dependence (as in Remark 4 above) but no logarithmic
dependence on $m^2$. This too is in contrast to the bound in [6].
These bounds for $d=2$ are thus stronger than  the estimate in Proposition 10.1 of
[6].

\vglue0.3cm

In the next two sections we will give proofs of the above
theorems. Before embarking on the proofs we indicate the strategy.
First we note that provided $m\neq 0$  the resolvent $G_{\a} (x-y,m^2)$ 
is in $L^{1}({\math Z}^{d})$ and
therefore periodizable (see e.g. Theorem 2.4 of Stein and Weiss [8],
Ch. 7, page 251) and the periodized version exists as an $L^{1}(T_{N+1})$ function on the
torus.  Provided $m\neq0$, the function $\tilde{\cal G}_{N,\a,  T_{N+1}}(\cdot,m^{2})$ is the
periodized version of an $L^{1}({\math Z}^{d})$ function $\tilde{\cal
G}_{N,\a}$ which we identified earlier. The periodized function is
in  $L^{1}(T_{N+1})$  and thus has a multiple Fourier
series. 
This is
obtained by a Poisson summation formula for the discrete
torus. Fourier analysis for finite Abelian groups is discussed in [8]. \color{black}
We will prove that
the Fourier coefficients, 
supplied by the discrete Fourier
transform, 
have rapid decay which leads not
only to the existence but also to very good uniform
differentiability properties of the periodic function. This is at the
heart of Theorem 1.1. The continuity
results of Theorem 1.2 will turn out to be relatively easy consequences. 

\vglue0.5cm
\\{\bf 2.  Proof of Theorem 1.1} 
\numsec=2\numfor=1

\\The function $G_{\a}(x-y,m^2)$ on ${\math Z}^{d}$ is pointwise
positive. This follows from the fact that it is the resolvent of
an $\a$- stable continuous time L\'evy walk $x_{t}^{(\a)} \in  {\math Z}^{d}$:

$$G_{\a}(x-y,m^2) = \int_{0}^{\infty} dt\> e^{-m^{2}t}\>
E_{x}(x^{(\a)}_{t}=y) . \Eq(2.1) $$
\\Therefore for $m\neq 0$ 

$$\eqalign {||G_{\a}(\cdot, m^{2})||_{L^{1}({\math Z}^{d})} &=  \int_{ {\math Z}^{d}} dx\>  |G_{\a}(x, m^{2})|      
= \int_{ {\math Z}^{d}} dx\>  G_{\a}(x, m^{2})\cr 
&=\hat G_{\a}(0, m^{2}) \cr                                    
&={1\over m^{2} }<\infty . \cr } \Eq(2.2)$$

\\Hence for $m \neq 0$ the function $G_{\a}(\cdot, m^{2})$ is in $L^{1}({\math
Z}^{d}) $ as claimed and the series 

$$G_{\a, T_{N+1}}(x, m^{2})= \sum_{{\ul n}\in {\math Z}^{d}}
\>G_{\a}(x +{\ul n}\> L^{N+1} , m^{2}) \Eq(2.3)$$

\\converges absolutely in the norm of $L^{1}(Q_{N+1})$ and defines a
function in  $L^{1}(T_{N+1}) $ (see [8], Ch.7, Theorem 2.4). 

\vglue0.3cm

\\{\it Remark}: Since $G_{\a}(\cdot, m^{2})$ is in $L^{1}({\math
Z}^{d}) $ for $m\neq 0$ it follows that there exists a $\d>0$ such that as
$|x|\rightarrow \infty$, $G_{\a}(x, m^{2})  \sim O(|x|^{-(d+\d)})$. In fact a precise estimate shows that $\d=\a$
where $0<\a<2$. However this fact will play no role in the rest of
this paper.

\vglue0.3cm

\\We now proceed to the proof of the finite range decomposition
\equ(1.2) and the bounds stated in Theorem 1.1. We only need to prove
the existence of the function  $\tilde{\cal G}_{N,\a,  T_{N+1}}  (x, m^{2})$ of \equ(1.6)
in $L^{1}(T_{N+1}) $ and the bound \equ(1.7). As in the proof of
Theorem 1.1 of [1, 2] given in section 3 of [1, 2], we will start with the finite
range decomposition of the resolvent of the Laplacian in ${\math Z}^{d}$ given in 
[3].  We will stop after the  first $N-1$ terms and then use the formula for the
remainder. From equation (3.31)  of [3]  we have 

$$G(x-y, s)=\sum_{j=0}^{N-1}\> L^{-j(d-2)}\>
\G_{j}(\bigl {(x-y\over L^{j}}, L^{2j}s \bigr )\> 
+ L^{-N(d-2)}\> {\cal G}_{N}\> \bigl
({x-y\over L^{N}}, L^{2N}s \bigr) \Eq(2.4)$$

\\$\G_j$ and ${\cal G}_N$ are defined through equations (3.28), (3.29)
and (3.30) of [3] (see equations (2.11), (2.12) and (2.13) below).
The products in these equations are convolution products. 

\\We now proceed as in Section 3 of [1], [2]. We insert the above finite
range decomposition with remainder in the Fourier transform of 
equation (2.2) of Proposition 2.1 of [1] to get (see equations (3.8)-
(3.12) of [1, 2]):

$$G_{\a} (x-y, m^{2})=\sum_{j=0}^{N-1}\> L^{-2j [\f]}\>
\G_{j,\a}(\bigl {(x-y\over L^{j}}, L^{j\a}m^{2} \bigr )\> 
+ L^{-2N[\f]}\> {\cal G}_{N,\a}\> \bigl
({x-y\over L^{N}}, L^{N\a}m^{2} \bigr) \Eq(2.41)$$

\\Because of their support properties (see \equ(1.4)), the periodization (with period
$L^{N+1}$) of the functions

$$\tilde \G_{j,\a}(x-y, m^{2})=L^{-2j[\f]} \G_{j,\a} \bigl ( {x-y\over L^{j}}, L^{j\a}m^{2}\bigr) \Eq(2.411)$$

\\for $0\le j \le N -1$ gives back the same functions which are 
therefore also defined on the torus $T_{N+1}$. 
Moreover Theorem 1.1 and Corollary 1.2 of [1] (corrected in the erratum)
and [2] gives the bounds \equ(1.14) and \equ(1.141)  on
these functions. It therefore remains to study
the torus boundary function

$$\tilde {\cal G}_{N,\a}(x-y)=     L^{-2N[\f]}\> {\cal G}_{N,\a}\> \bigl
({x-y\over L^{N}}, L^{N\a}m^{2} \bigr) \Eq(2.5)$$

\\and its periodization. Now

$${\cal G}_{N,\a}(\cdot,m^{2}): (\e_N{\math Z})^{d}\rightarrow {\math R} \Eq(2.6)$$

\\is given by 

$${\cal G}_{N,\a}(\cdot,m^{2}) = \int_{0}^{\infty}ds\>\r_{\a}(s,\>m^{2})\> 
{\cal G}_{N} (\cdot,\>s )\Eq(2.7)$$

\\where $\r_{a}$ is the spectral function of Proposition 2.1 of
[1]. Let us introduce the notation

$${\cal G}_{N} (s )(x-y)=  {\cal G}_{N} (x-y,\>s ) . \Eq(2.8)$$

\\{\it Claim}: 

$${\cal G}_{N} (s )(x-y) \ge 0 \Eq(2.81)$$

\\{\it Proof}

\\From equations (3.28) and (3.30) of [4] we have

$${\cal G}_{N} (s )={\cal A}_{N}(s)G_{\e_N}(s) {\cal A}_{N}(s)^{*} . \Eq(2.9)$$

\\ $ G_{\e_N}(s)(u-v)$ is the
resolvent of the laplacian on the lattice $(\e_N{\math Z})^{d}$ 
and the products in \equ(2.9) are  convolution products with
(defective) probability measures:

$${\cal G}_{N} (s )(x-y) =\int_{(\e_N{\math Z})^{d}} \int_{(\e_N{\math
Z})^{d}}  {\cal A}_{N}(s)(x,du)G_{\e_N}(s)(u-v) {\cal A}_{N}(s)(y,dv)
\Eq(2.10)$$

\\and ${\cal A}_{N}(s)$ is given by a convolution product of averaging operators:

$$ {\cal A}_{N} (s)=\prod_{m=1}^{N} A_{\e_j, m}
(L^{-(m-1})(s) \Eq(2.11)$$

\\which themselves are (defective) probability measures. A probability
measure is called defective if its total mass is less than $1$, which
is the case if $s>0$. Now the action of each averaging operator on a
function $f$ is given by equation (3.23) of [4]. It is composed of a
non-negative constraining function and the action of a Poisson kernel
measure whose action is positivity preserving. Therefore the action of
each averaging operator is positivity preserving and hence their
convolution product ${\cal A}_{N}(s)$ is positivity
preserving. Finally $G_{\e_N}(s)(u-v)$ being the resolvent of a random
walk in $(\e_N{\math Z})^{d}$ is pointwise positive. Therefore  
${\cal G}_{N} (s )(x-y) \ge 0$. \bull 

\\Therefore

$$\eqalign {||{\cal G}_{N}(s)||_{L^{1}(({\e_N\math Z})^{d})} &=  \int_{
{(\e_N\math Z})^{d}} dx\>  |{\cal G}_{N}(x, s)|      
= \int_{(\e_N{\math Z})^{d}} dx\>   {\cal G}_{N}(x, s)  \cr 
&=\hat{\cal G}_{N}(0, s)  \cr} \Eq(2.18)$$

\\Now taking the Fourier transform of \equ(2.10) and \equ(2.11) we get

$$\hat{\cal G}_{N}(p, s)={|\hat{\cal A}_{N}(p,s)|^{2} \over s-\hat\D_{\e_N}(p)}  \Eq(2.19) $$

\\where $p\in B_{\e_N} =[-{\p\over \e_N} , {\p\over \e_N}]$ and
$\hat\D_{\e_N}(p)$ is the Fourier transform of the $\e_N$-lattice
Laplacian. From Appendix B of [2]  we have for every
integer $k\ge 0$ and $N\ge 2$ the bound

$$|\hat{\cal A}_{N}(p,s)|^{2} \le c_{L,k} (1+s)^{-2} (p^{2} +1)^{-k}
. \Eq(2.20)$$

\\From \equ(2.19) and
\equ(2.20) we get

$$|\hat{\cal G}_{N}(p, s)|\le c_{L, k} (1+s)^{-2} (p^{2} +1)^{-k}
(s-\hat\D_{\e_N}(p))^{-1} \Eq(2.21)$$

\\where $c_{L, k}$ is independent of $N$. 
Therefore for $s>0$ we have from \equ(2.18), \equ(2.19) and \equ(2.21) 

$$||{\cal G}_{N}(s)||_{L^{1}(({\e_N\math Z})^{d})} \le c_{L, k} 
(1+s)^{-2} \> {1\over s}  . \Eq( 2.22)$$

\\From \equ(2.7), \equ(2.21) and the bound on the spectral function
$\r(s,m^{2}) $ in Proposition 2.1, equation (2.4) of [1], [2] 

$$0\le \r_{\a} (s, m^{2})\le c_{\a}\> {s^{\a/2} \over s^{\a} + m^{4}} 
\Eq(2.222) $$

\\and hence we get for $m\neq 0$

$$\eqalign{ ||{\cal G}_{N,\a}(\cdot,m^{2})||_{L^{1}(({\e_N\math Z})^{d})}  &\le c_{\a}
\int_{0}^{\infty}ds\> {s^{\a/2} \over s^{\a} + m^{4}}   \> 
||{\cal G}_{N} (\cdot,\>s )||_{L^{1}(({\e_N\math Z})^{d})} \cr
&\le c_{\a, L, k} \int_{0}^{\infty}ds\> {s^{\a/2 -1} \over s^{\a} + m^{4}} (1+s)^{-2}\cr
&\le c_{\a, L, k}\> c_{\a} {1\over m^{4}} . \cr} \Eq(2.23)   $$

\\where the constant

$$c_{\a} =\int_{0}^{\infty}ds\> s^{\a/2 -1}  (1+s)^{-2} <\infty \Eq(2.24)$$

\\is O(1) since $0<\a<2$. It follows from \equ(2.5) and the bound in
\equ(2.23) that for $m\neq 0$

$$\eqalign{||\tilde{\cal G}_{N,\a}(\cdot,m^{2})||_{L^{1}({\math Z}^{d})} 
&=L^{N\a} ||{\cal G}_{N,\a}(\cdot,L^{N\a}m^{2})||_{L^{1}({(\e_N\math Z})^{d})}\cr
&\le c_{L,\a,k}c_{\a} L^{N\a}  {1\over L^{2N\a}m^{4}} . \cr}\Eq(2.25)$$

\\Therefore for $m\neq 0$, $\tilde{\cal G}_{N,\a}(\cdot,m^{2})$ is in $L^{1}({\math Z}^{d})$
and

$$\tilde{\cal G}_{N,\a, T_{N+1}}(x,m^{2})= \sum_{y \in {\math
L^{N+1}{\math Z}^{d}} }\tilde{\cal G}_{N,\a}(x +y ,m^{2})  \Eq(2.26)$$

\\converges absolutely in the norm of $L^{1}(Q_{N+1}) $ and hence
defines a function in $L^{1}(T_{N+1}) $.
As a $L^{1}(T_{N+1})$   periodic function
$\tilde{\cal G}_{N,\a, T_{N+1}}(\cdot,m^{2})$  has a Fourier series
which is obtained by Poisson summation with discrete Fourier transform
for the discretized torus:

$$\tilde{\cal G}_{N,\a, T_{N+1}}(x,m^{2})= {1\over |Q_{N+1}|}
\sum_{p\in {2\p\over L^{N+1}}  Q_{N+1} } \hat{\tilde{\cal G}}_{N,\a}(p,m^{2})
 e^{ i p.x}  . \Eq(2.27)$$ 

\\Recall that $Q_{N+1}= [-{L^{N+1}\over 2}, {L^{N+1}\over 2}]^{d}\cap  {\math
Z}^{d}$. Here $L=3^{p}$ where $p\ge 1 $ is any positive integer (this
$p$ is not be confused with the $p$ appearing in the sum).
Thus the sum is over a discretization of the Brillouin zone
 $B_{\e_{0}} =[-\p, \p] ^{d}$ where the discrete
 Fourier transform in ${\math Z}^{d} $ occuring in \equ(2.27) is defined. We shall now estimate the
 Fourier coefficients.

\vglue0.3cm

\\From the definition 

$$ \tilde{\cal G}_{N,\a, }  (x, m^{2})=L^{-2N[\f]}
 {\cal G}_{N,\a}  ({x\over L^{N}}, L^{N\a}m^{2}) \Eq(2.271)$$

\\we obtain for the Fourier transform

$$\eqalign{\hat{\tilde{\cal G}}_{N,\a}(p,m^{2})&=L^{N\a} \hat{{\cal
G}}_{N,\a}(L^{N} p, L^{N\a}m^{2}) \cr
&= L^{N\a}\int_{0}^{\infty} ds\> \r_{\a} (s, L^{N\a} m^{2}) \hat{\cal
G}_N(L^{N}p,s)  \cr }  \Eq(2.272)$$

\\where $p\in [-\p, \p]^{d}$ and  $\r_{\a}$ is that of \equ(1.153). 
Using the bounds supplied in \equ(2.21)
and \equ(2.222) we obtain for $ m\neq 0$

$$\eqalign{ |\hat{\tilde{\cal G}}_{N,\a}(p,m^{2})| &\le c_{\a} c_{L,k}L^{N\a} 
\int_{0}^{\infty}ds\> {s^{\a/2-1} \over s^{\a} + L^{2N\a}m^{4}} (1+s)^{-2}  \> 
((L^{N}p)^{2} +1)^{-k}\cr
&\le c_{\a, L, k} {L^{N\a} \over (L^{N\a}m^{2})^{2}}
((L^{N}p)^{2}+1)^{-k}  . \cr} \Eq(2.273)  $$

\\From \equ(2.27) we get

$$\tilde{\cal G}_{N,\a, T_{N+1}}(x,m^{2})= {1\over |Q_{N+1}|}
\sum_{p\in Q_{N+1} } \hat{\tilde{\cal G}}_{N,\a}({2\p\over L^{N+1}} p,m^{2})
 e^{ i {2\p\over L^{N+1}}  p.x} . \Eq(2.274)$$ 

\\Taking partial derivatives of order $l$ with respect to  $x\in {\math Z}^{d}$ we get 

$$\dpr_{{\math Z}^{d}} ^{l}\> \tilde{\cal G}_{N,\a,T_{N+1}}(x,m^{2})|={1\over |Q_{N+1}|}
\sum_{p\in Q_{N+1} }  ({2\p p\over L^{N+1}})^{l}  \hat{\tilde{\cal G}}_{N,\a}({2\p\over L^{N+1}} p,m^{2})
 e^{ i {2\p\over L^{N+1}}  p.x}  \Eq(2.275)$$ 

\\where the partial derivative $\dpr_{{\math Z}^{d}} ^{l}$ is in
 multi-index notation, 
$p^{l}= \prod _{i=1} ^{d} p_{i}^{l_i} $, $l_i\ge 0$ are
 non-negative integers and $l=\sum_{i=1}^{d} l_i \> \ge 0$. We now use
 the bound \equ(2.273) and extend the sum to ${\math Z}^{d}$ to get

$$|\dpr_{{\math Z}^{d}} ^{l} \> \tilde{\cal
G}_{N,\a,T_{N+1}}(x,m^{2})|\le   c_{\a, L, k}  L^{-(N+1)d} {L^{N\a} \over (L^{N\a}m^{2})^{2}}
\sum_{p\in {\math Z}^{d } } ({2\p |p|\over L^{N+1}})^{l}  ({(2\p
p\over L})^{2}+1)^{-k} 
 . \Eq(2.276)$$ 

\\The non-negative integer $k$ is at our disposal. 
We choose $2k>d+l+1$.  Then the series converges and we get the bound for all
$m\neq 0$ and all integers $l\ge 0$ 

$$|\dpr_{{\math Z}^{d}} ^{l}\> \tilde{\cal G}_{N,\a,T_{N+1}}(x,m^{2})|
\le c_{L,\a, l}  L^{-2N\a} m^{-4} \>  L^{-(2N[\f] +lN)}
. \Eq(2.36)$$

\\This proves \equ(1.7) and thus the proof of Theorem 1.1 is
complete. \bull 

\vglue0.3cm

\\The next section is devoted to the proof of Theorem 1.2

\vglue0.5cm
\\{\bf 3.  Proof of Theorem 1.2} 
\numsec=3\numfor=1
\vglue0.3cm
\\Throughout the proof we restrict $\a$ to the range $1<\a<2$.

\vglue0.3cm
\\First, we will prove the differentiability bound. Recall the rescaled fluctuation covariances  

$$\G_{j,\a} (\cdot,\>m^{2} ) :\>(\e_{j} {\math Z} )^{d} \rightarrow {\math R}$$

\\defined in equation (3.11) in Section 3 of [1, 2].

$$ \G_{j,\a} (\cdot,\>m^{2} )= \int_{0}^{\infty} ds\>\r_{\a}(s,\>
m^{2})\> \G_{j} (\cdot,\>s ) $$

\\where $\G_{j} (\cdot,\>s )$ is the rescaled fluctuation covariance
in the finite range decomposition of the resolvent of the standard
Laplacian (see Section 3 of [1, 3]).
Then we have for all $0\le j \le N-1$, on using the uniform bound in 
Theorem 5.5 of [4] together with Sobolev embedding:

$$\eqalign{|| {\dpr\over \dpr m^{2}} \dpr_{\e_j}^{p}\G_{j,\a} (\cdot,\>m^{2} )||_{L^{\infty}
((\e_q{\math Z})^{d})} &\le \int_{0}^{\infty} ds\>
|{\dpr\over \dpr m^{2}}\r_{\a}(s,\>  m^{2})|\> ||\dpr_{\e_j}^{p}\G_{j} (\cdot,\>s )||_{L^{\infty}
((\e_q{\math Z})^{d})}\cr
&\le c_{L, p} \>\int_{0}^{\infty} ds\>
|{\dpr\over \dpr m^{2}}\r_{\a}(s,\>  m^{2})| (1+s)^{-1}  \cr  } \Eq(3.111)$$

\\where $\r_{a}(s, m^{2})$ is the spectral function of
Proposition 2.1 of [1]:

$$\r_{\a} (s, m^{2}) = {\sin\p\a/2\over \p} \> {s^{\a/2}\over s^{\a}
+m^{4} + 2 m^{2} s^{\a/2} \cos\p\a/2} \> . \Eq(3.114)    $$

\\Hence

$$\eqalign{|{\dpr\over \dpr m^{2}}\r_{\a} (s, m^{2})| &\le c_{\a}
{s^{\a/2} (m^{2} + s^{\a/2})   \over (s^{\a}
+m^{4} + 2 m^{2} s^{\a/2} \cos\p\a/2)^{2} } \cr
&\le c_{\a} \>{s^{\a/2} (m^{2} + s^{\a/2})   \over (s^{\a}
+m^{4}  )^{2} } \cr}\Eq(3.115)    $$

\\where we have used from the proof of Proposition 2.1 of [1] the bound

$$\eqalign{d_{\a}(s, m^{2}) &= s^{\a}+m^{4} + 2 m^{2}\> s^{\a/2}
\cos\p\a/2 \cr
&\ge c'_{\a}\>(m^{4} + s^{\a}) . \cr} $$

\\Therefore
$$||{\dpr\over \dpr m^{2}} \dpr_{\e_j}^{p}\G_{j,\a} (\cdot,\>m^{2} )||_{L^{\infty}
((\e_q{\math Z})^{d})}
\le  c_{L, \a, p}\int_{0}^{\infty} ds\>
 {s^{\a/2} (m^{2}+s^{\a/2}) \over (s^{\a} + m^{4})^{2}}  (1+s)^{-1} \>
 . \Eq(3.112)$$

\\ After a change of variables $s^{\a/2} =m^{2}\s$ we get with a
 different constant $c_{L\a, p}$

$$ ||{\dpr\over \dpr m^{2}} \dpr_{\e_j}^{p}\G_{j,\a} (\cdot,\>m^{2} )||_{L^{\infty}
((\e_q{\math Z})^{d})}   \le c_{L, \a, p} 
(m^{2})^{{2\over\a}-2} H_{\a} (\m)  \Eq(3.113) $$

\\where 

$$\m= (m^{2})^{2\over\a} \Eq(3.7)$$ 

\\and 

$$ H_{\a} (\m)= \int_{0}^{\infty}d\s \>   {{\s}^{2\over\a}
(1+{\s}) \over (1+ {\s}^{2} )^{2}} 
(1+\m \s^{2\over \a} )^{-1}
 . \Eq(3.8)$$

\\For $1<\a<2$ we have the obvious bound

$$H_{\a} (\m) \le  H_{\a} (0) \Eq(3.81)$$
where

$$H_{\a} (0)= \int_{0}^{\infty}d\s \>   {{\s}^{2\over\a}
(1+{\s}) \over (1+ {\s}^{2} )^{2}}  <\infty   \Eq(3.9)$$

\\since $1<\a<2$ and thus $H_{\a} (0)$ is a constant $c_{\a}$ of
$O(1)$. 

\\From \equ(3.113), \equ(3.7), \equ(3.8), \equ(3.81) and \equ(3.9) we get

$$||{\dpr\over \dpr m^{2}} \dpr_{\e_j}^{p}\G_{j,\a} (\cdot,\>m^{2} )||_{L^{\infty}
((\e_q{\math Z})^{d})}   \le c_{L, \a, p} 
(m^{2})^{{2\over \a} -2}   \Eq(3.11)$$ 

\\and hence 

$$\eqalign{||{\dpr \over \dpr m^{2}} \dpr_{{\math Z} ^{d}}^{p}\tilde\G_{j,\a}(\cdot, m^{2})
||_{L^{\infty} ({\math Z}^{d})} &\le c_{L, \a, p} L^{-pj} L^{-2j[\f]} L^{j\a} ( L^{j\a}
  m^{2})^{-2(1-{1\over \a})} \cr  &= c_{L, \a, p}L^{-j(d-2)} L^{-pj}
  (m^{2})^{-2(1-{1\over \a})}   \cr}\Eq(3.121)$$

\\which proves the differentiability bound \equ(1.101) of Theorem
  1.2.  The uniform continuity bound \equ(1.103) now follows by integrating
the bound \equ(3.121) above. 

\vglue0.3cm

\vglue0.3cm

\\It remains now to prove the uniform Lipshitz continuity bound
\equ(1.102) of $\tilde{\cal G}_{N,\a,  T_{N+1}}  (\cdot,
m^{2})$. Recall that in the uniform continuity statement
$m^{2}>0$. We
will first give an uniform upper bound for its derivative with respect
to $m^{2}$ from which the uniform Lipshitz continuity bound will follow. To
this end
we start from the Fourier series representation \equ(2.275) where the
Fourier coefficients decay rapidly as in \equ(2.273). We shall show
presently that their derivatives with respect to $m^{2}$ also decay rapidly.
We have

$$||{\dpr\over \dpr m^{2}}\dpr_{{\math Z}^{d}} ^{l}\> \tilde{\cal
G}_{N,\a,T_{N+1}}(\cdot,m^{2})||_{L^{\infty} ({\math Z}^{d})} \le {1\over |Q_{N+1}|}
\sum_{p\in Q_{N+1} }  ({2\p p\over L^{N+1}})^{l} |{\dpr\over \dpr m^{2}}\hat{\tilde{\cal G}}_{N,\a}({2\p\over L^{N+1}} p,m^{2})|
 . \Eq(2.285)$$

\\Now using the representation \equ(2.7), the equality \equ(2.272),
and the bound \equ(2.21)  we
obtain for every integer $k\ge 0$

$$\eqalign{\bigl
|{\dpr\over \dpr m^{2}}\hat{\tilde{\cal G}}_{N,\a}({2\p\over L^{N+1}} p,m^{2})|  \bigr| 
&\le c_{L,k} L^{2N\a} \int_{0}^{\infty} ds\>
\bigl| {\dpr\over \dpr m^{2}}\r_{\a}(s, m^{2})
\bigr|_{m^{2}\rightarrow L^{N\a} m^{2}}  \times\cr
&  (1+s)^{-2}
\bigl(({2\pi p\over L})^{2}  +1\bigr)^{-k} s^{-1} . \cr} \Eq(3.13)$$

\\We have the bound (this was obtained in going from \equ(3.111) to \equ(3.112))

$$\bigl| {\dpr\over \dpr m^{2}}\r_{\a}(s, m^{2})
\bigr| \le c_{\a} {s^{\a/2} (m^{2}+s^{\a/2}) \over (s^{\a} +
m^{4})^{2}} .\Eq(3.14)$$

\\From \equ(2.285), \equ(3.13) and \equ(3.14) and then extending the
sum to that over ${\math Z}^{d}$  we obtain by choosing $k$
sufficiently large so that the series converges,

$$\eqalign{||{\dpr\over \dpr m^{2}}\dpr_{{\math Z} ^{d}}^{l} \tilde{\cal G}_{N,\a,  T_{N+1}}  (\cdot, m^{2})||_{L^{\infty} ({\math Z}^{d})}
&\le   c_{L,\a, l}L^{-(N+1)d}L^{-Nl}  L^{2N\a} \times\cr
&\int_{0}^{\infty} ds\>{s^{\a/2 -1} (m^{2}+s^{\a/2}) \over (s^{\a} +(m^{2})^{2})^{2}}
(1+s)^{-2}  \bigr|_{m^{2}\rightarrow L^{N\a} m^{2}} \cr} \Eq(3.15)$$

\\where the integral

$$F_{\a}(m^{2}) =\int_{0}^{\infty} ds\>{s^{\a/2 -1} (m^{2}+s^{\a/2}) \over (s^{\a} +(m^{2})^{2})^{2}}
 (1+s)^{-2} \Eq(3.16)$$

\\converges since $\a>0 $ and $m\neq 0$. We now change variables
as in the line before \equ(3.113): $s^{\a/2} =m^{2}\s$ to get  with $\m= (m^{2})^{2\over\a}$

$$\eqalign{F_{\a}(m^{2})&= (m^{2})^{-2} \>{2\over \a}\int_{0}^{\infty}d\s \>  
{(1+{\s}) \over (1+ {\s}^{2} )^{2}} 
(1+\m \s^{2/\a})^{-2}\cr
&\le (m^{2})^{-2}\>{2\over \a} \int_{0}^{\infty}d\s \>  
{(1+{\s}) \over (1+ {\s}^{2} )^{2}}\cr
&\le c_{\a}\>  (m^{2})^{-2}  \cr}   \Eq(3.17)$$ 

\\since the last integral on the right converges to a constant of O(1).
Therefore 

$$F_{\a}(L^{N\a}m^{2})\le L^{-2N\a}\>c_{\a} (m^{2})^{-2}  . \Eq(3.18)  $$ 

\\From \equ(3.15), \equ(3.16) and \equ(3.18)
we get for all integers $l\ge 0$

$$||{\dpr\over \dpr m^{2}}\dpr_{{\math Z} ^{d}}^{l}\tilde{\cal G}_{N,\a,  T_{N+1}}  (\cdot, m^{2})||_{L^{\infty} ({\math Z}^{d})}
\le   c_{L,\a, l}\>L^{-(N+1)d} L^{-Nl}  (m^{2})^{-2} .\Eq(3.19)$$
\\The uniform continuity bound for non-zero mass is now obtained by integration. The
proof of Theorem 1.2 is complete. \bull

\vglue0.3cm
\\{\bf Appendix A}

\vglue0.3cm

\\In this Appendix we prove the statements in the first paragraph of Remark 4. By definition
the fluctuation covariances on the coarser scale $L'= L^{r}$ with
$L\ge 2$
fixed and $r$ a large positive integer is given by \equ(1.1411):

$$\tilde \G'_{j,\a} (\cdot,m^{2})= \sum_{l=0}^{r-1} \tilde \G_{l+jr,\a} (\cdot,
m^{2})  . \Eq(1.1413)$$ 

\\Therefore we get

$$\eqalign {||{\dpr \over \dpr m^{2}} \dpr_{ {\math Z}^{d} }^{p} \tilde\G'_{j,\a}(\cdot, m^{2})
||_{L^{\infty} ({\math Z}^{d})} &\le  \sum_{l=0}^{r-1} ||{\dpr \over \dpr m^{2}} \dpr_{ {\math Z}^{d} }^{p} \tilde \G_{l+jr,\a} (\cdot,
m^{2}) ||_{L^{\infty} ({\math Z}^{d})} \cr
&\le c_{L, \a, p}  (m^{2})^{-2(1-{1\over \a})}    \sum_{l=0}^{r-1}   L^{-p(l+jr)} L^{-(l+jr)(d-2)} \cr
&\le c_{L, \a, p}  (m^{2})^{-2(1-{1\over \a})} (L')^{-pj}
(L')^{(d-2)j}   \sum_{l=0}^{\infty} L^{-(p +(d-2))l} . \cr} $$

\\For $d\ge 3$, $\forall p \ge 0$ and $d=2$, $\forall p \ge 1$ we can bound the sum on the right hand side by

$$\sum_{l=0}^{\infty} L^{-l}   = (1-{1\over L} )^{-1}       $$

\\and hence

$$||{\dpr \over \dpr m^{2}} \dpr_{ {\math Z}^{d} }^{p} \tilde\G'_{j,\a}(\cdot, m^{2})
||_{L^{\infty} ({\math Z}^{d})} \le c'_{L, \a, p}  (m^{2})^{-2(1-{1\over \a})} (L')^{-pj}
(L')^{(d-2)j} . $$

\\which is \equ(1.102) with a new constant independent of $L'$ as
claimed. \bull

\vglue0.3cm
\\{\bf Acknowledgements} I wish to thank David Brydges for many helpful
conversations and for setting me right on Poisson summation for a
discrete torus.  I also thank the diligent reviewers for their remarks,
questions and suggestions.

\vglue0.5cm

\\{\bf References}

\vglue0.3cm
\\[1] P. K. Mitter: On a finite range decomposition of the resolvent
of a fractional power of the laplacian,
J Stat Phys (2016) {\bf 163}:1235-1246, Erratum: J Stat Phys (2017)
{\bf 166} : 453-455

\vglue.3cm
\\[2] P. K. Mitter: On A Finite Range Decomposition of the Resolvent
of a Fractional Power of the Laplacian: http://arxiv.org/abs/1512.02877

\vglue.3truecm  
\\[3] D. Brydges, G. Guadagni and P. K. Mitter: Finite range
Decomposition of Gaussian Processes,
J. Stat. Phys. (2004) {\bf 115}: 415--449  

\vglue0.3cm
\\[4] Roland Bauerschmidt: A simple method for finite range decomposition
of quadratic forms and Gaussian fields,
Probab. Theory Relat. Fields (2013) {\bf 157}: 817-845

\vglue0.3cm
\\[5] Roland Bauerschmidt (unpublished)
\vglue.3cm

\vglue0.3cm
\\[6]  G. Slade: Critical exponents for long range $O(n)$ models below the
upper critical dimension, https://arxiv.org/1611.06169

\vglue.3truecm
\\[7] Elias M Stein and Guido Weiss: Introduction to Fourier Anaysis
on Euclidean Spaces,
Princeton University Press, second printing (1975), Princeton, New Jersey

\vglue0.3cm
\\[8] Audrey Terras: Fourier Analysis on Finite Groups and Applications, London
Mathematical Society Student Texts 43, Cambridge University Press,
second printing (2001), Cambridge UK.

\vglue0.3cm
\\[9] Stefan Adams, Roman Koteck\'y, Stefan Muller: Finite range
decomposition for families of gardient Gaussian measures, 
J. Funct. Anal. (2013) {\bf 264}: 169-206

\vglue0.3cm
\\[10] R. Bauerschmidt, D. C. Brydges, G. Slade: A renormalisation
group method. III. Perturbative analysis,
J Stat Phys (2015) {\bf 159}: 492-529

\bye